\documentclass[final,5p,times,twocolumn]{elsarticle}

\usepackage{graphicx}

\usepackage{amssymb}

\usepackage{wasysym}

\journal{Journal of Physics and Chemistry of Solids}

\begin{document}

\begin{frontmatter}



\title{Optical properties of the iron-chalcogenide superconductor
  FeTe$_{0.55}$Se$_{0.45}$}

%
\author[]{Christopher C. Homes\corref{cor1}}
\ead{homes@bnl.gov}
\author{Ana Akrap}
\author{Jinsheng Wen}
\author{Zhijun Xu}
\author{Zhi Wei Lin}
\author{Qiang Li}
\author{Genda Gu}

\address{Condensed Matter Physics and Materials Science Department,
  Brookhaven National Laboratory, Upton, New York 11973, USA }
\cortext[cor1]{Corresponding author.}

\begin{abstract}
The complex optical properties of the iron-chalcogenide superconductor
FeTe$_{0.55}$Se$_{0.45}$ with $T_c = 14$~K have been examined over a wide
frequency range for light polarized in the Fe-Te(Se) planes above and below
$T_c$.  At room temperature the optical response may be described by a
weakly-interacting Fermi liquid; however, just above $T_c$ this picture breaks
down and the scattering rate takes on a linear frequency dependence.
Below $T_c$ there is evidence for two gap features in the optical
conductivity at $\Delta_1 \simeq 2.5$~meV and $\Delta_2 \simeq 5.1$~meV.
Less than 20\% of the free carriers collapse into the condensate for
$T\ll T_c$, and this material is observed to fall on the universal
scaling line for a BCS dirty-limit superconductor in the weak-coupling limit.

\end{abstract}


\end{frontmatter}



\section{Introduction}
\label{}

There have been many surprises in the field of superconductivity in the last
25 years.  First and foremost was the discovery of superconductivity at
elevated temperatures in the copper-oxide materials \cite{bednorz86}.  For
conventional metals and alloys, in the model developed by Bardeen, Cooper
and Schrieffer (BCS) superconductivity is mediated through lattice
vibrations where electrons form bound pairs \cite{bcs}; the condensation
below the critical temperature ($T_c$) is also accompanied by the formation
of an isotropic {\em s}-wave energy gap at the Fermi surface.
Within this framework, it was thought that the critical temperature could not
exceed approximately 30~K \cite{mcmillan68}.  With $T_c$'s in excess of 130~K at
ambient pressure and an unusual {\em d}-wave energy gap with nodes at the
Fermi surface, the pairing mechanism in the cuprates remains unresolved.
The discovery of superconductivity in MgB$_2$ with the surprisingly high
transition temperature of $T_c = 39$~K \cite{nagamatsu01} initially suggested
an unusual pairing mechanism; however, the isotope effect established that the
superconductivity in this material is likely phonon mediated \cite{budko01}.
In this particular case, the high phonon frequencies in MgB$_2$ are
likely responsible for the unusually large value for $T_c$ \cite{kortus01}.

\begin{figure}[htbp]
\centerline{\includegraphics[width=8.0cm]{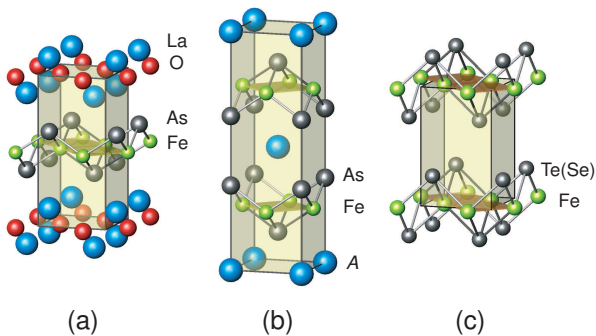}}
\caption{The tetragonal extended unit cells of the iron-pnictide materials
 (a) LaFeAsO, (b) $A$Fe$_2$As$_2$, where $A$ is an alkali earth, and (c) the
 iron-chalcogenide FeTe.  In the first two materials, the iron-arsenic
 sheets are separated by LaO or alkali-earth layers, respectively.
 However, the iron-chalcogenide material is structurally simpler,
 consisting only of FeTe(Se) layers.}\label{fig:unitcell}
\end{figure}

The discovery of superconductivity in the iron-arsenic LaFeAsO$_{1-x}$F$_x$
(1111) pnictide compound \cite{kamihara08} was surprising because iron had
long been considered detrimental to superconductivity.  Different rare earth
substitutions in this material quickly raised $T_c \gtrsim 50$~K
\cite{ren08a,ren08b,ishida09}.  While such high values for $T_c$ do not definitively
rule out a phonon-mediated pairing mechanism, the presence of magnetic order
close to the superconductivity in these compounds \cite{cruz08} has led to
the suggestion that the pairing in this class of materials may have
another origin \cite{boeri08,johnston10}.    The highest $T_c$'s have been observed
in the 1111-family of materials; however, large single crystals have only
recently been obtained \cite{yan09}; the extended unit cell of
non-superconducting LaFeAsO is shown in Fig.~\ref{fig:unitcell}(a).
As a consequence, much of the focus has shifted to the structurally-simpler
$A$Fe$_2$As$_2$ (122) iron-pnictides, shown in Fig.~\ref{fig:unitcell}(b),
and the FeTe(Se) (11) iron-chalcogenide materials, shown in Fig.~\ref{fig:unitcell}(c),
where large single crystals are available.  The metallic $A$Fe$_2$As$_2$
materials (where $A=$ Ca, Ba or Sr) have been extensively studied; in
BaFe$_2$As$_2$ the application of pressure results in $T_c \simeq 29$~K, while
Co- and Ni-doping yields $T_c \simeq 23$~K at ambient pressure
\cite{alireza09,sefat08,li09}.
Superconductivity has been observed in the arsenic-free iron-chalcogenide
FeSe compound with $T_c = 8$~K, which increases to $T_c \simeq 37$~K under
pressure \cite{hsu08,mizuguchi08,medvedev09,margadonna09}.  Through the
substitution of Se for Te the critical temperature at ambient pressure
reaches a maximum $T_c = 14$~K in FeTe$_{0.55}$Se$_{0.45}$.  Despite the
structural differences of the iron-pnictides and the iron-chalcogenides
illustrated in Fig.~\ref{fig:unitcell}, the band structure of these materials
is remarkably similar, with a minimal description consisting of an electron
band $(\beta)$ at the $M$ point and a hole band $(\alpha)$ centered at the
$\Gamma$ point of the Brillouin zone \cite{raghu08}.

There have been a number of studies of the Fe$_{1+x}$Te and FeTe$_{1-x}$Se$_x$
materials, including transport \cite{fang08,sales09,taen09,chengf09},
tunneling \cite{kato09}, and angle-resolved photoemission (ARPES)
\cite{xiay09,nakayama09}, with particular attention on the magnetic
properties \cite{nakayama09,leesh09,wen09,qui09,khasanov09,han09,bao09,mook10}.
While the optical properties of the superconducting iron-pnictides
have been investigated in considerable detail
\cite{hu08,yang09,wu09a,kim09,heumen09,gorshunov10,perucchi10,wu10,nakajima10}, the
iron-chalcogenide materials are relatively unexplored \cite{chengf09}.

%
%

In this work we examine the in-plane complex optical properties of a
single crystal of superconducting FeTe$_{0.55}$Se$_{0.45}$ above and below
$T_c$.  At room temperature this material may be described as a weakly-interacting
Fermi liquid and the transport is Drude-like.  However, just above
$T_c$ this interpretation breaks down and the scattering rate adopts
a linear frequency dependence.  Below $T_c$ there are clear signatures of the
superconductivity in the reflectance and the optical conductivity.  Less than
20\% of the free carriers collapse into the superconducting condensate,
suggesting that this material is in the dirty limit, and this material
is observed to fall on the scaling line predicted for a BCS dirty-limit
superconductor in the weak-coupling limit.  In addition, there is evidence
for two gap features in at $\Delta_1 \simeq 2.5$~meV and $\Delta_2
\simeq 5.1$~meV.  Some of these results have been discussed in a previous
work \cite{homes10a}.

%
%
\section{Results and Discussion}
\begin{figure}[t]
\centerline{\includegraphics[width=7.3cm]{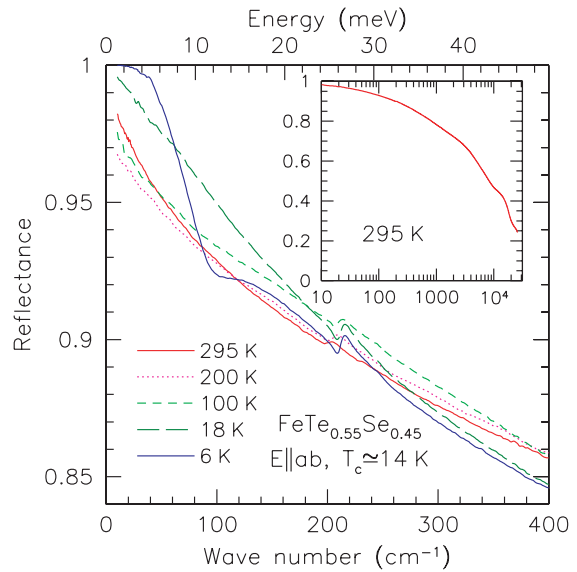}}
\caption{The temperature dependence of the reflectance of a cleaved surface
of FeTe$_{0.55}$Se$_{0.45}$ in the terahertz and far-infrared region
for light polarized in the {\em a-b} planes at several temperatures
above and below $T_c$.  There is a dramatic change in the reflectance below $T_c$.
Inset: The reflectance at room temperature over a wide frequency range.}
\label{fig:reflec}
\end{figure}

Single crystals with good cleavage planes (001) were grown by a unidirectional
solidification method with a nominal composition of FeTe$_{0.55}$Se$_{0.45}$.
The normal-state resistivity is in good agreement with literature values \cite{sales09}.
The critical temperature determined from magnetic susceptibility is $T_c = 14$~K
with a transition width of $\lesssim 1$~K.  The temperature dependence of the reflectance
has been measured at a near-normal angle of incidence on a freshly-cleaved surface above
and below $T_c$ over a wide frequency\footnote{Some useful conversions used in
this text are 1~eV = 8066~cm$^{-1}$, 1~THz = 33.4~cm$^{-1}$,  1~K = 0.695~cm$^{-1}$, and
1~$\Omega^{-1}$cm$^{-1}$ = 4.78~cm$^{-1}$.}
range ($\simeq 2$~meV to 3.5 eV) for light polarized
in the {\em a-b} planes using an {\em in situ} overcoating technique \cite{homes93}.
The reflectance in the terahertz and far-infrared region is shown in Fig.~\ref{fig:reflec};
the reflectance at 295~K is shown over a wider region in the inset.  At room
temperature, the reflectance displays the $R\propto 1 - \sqrt{\omega}$ response
characteristic of a metal in the Hagen-Rubens regime; however, at low temperature just
above above $T_c$ (18~K), $R \propto 1 - \omega$, which is reminiscent of the
reflectance in a marginal Fermi liquid \cite{varma89,hwang04}.  The development of the
superconducting state has a clear signature in the reflectance.  However, the
reflectance is a complex quantity consisting of an amplitude and a phase,
$\tilde{r} = \sqrt{R} e^{i\theta}$.  Normally, only the amplitude $R = \tilde{r}
\tilde{r}^\ast$ is measured so it is not intuitively obvious what changes in
the reflectance imply.  For this reason, the complex optical properties have
been calculated from a Kramers-Kronig analysis of the reflectance \cite{dressel-book}.

%
%

\begin{figure}[tb]
\centerline{\includegraphics[width=7.3cm]{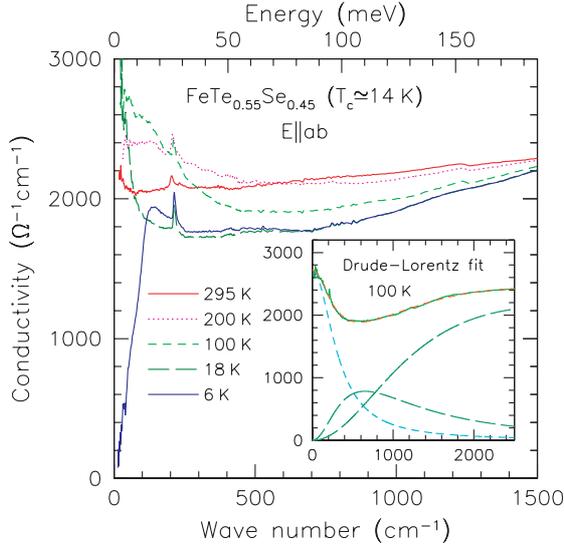}}
\caption{The real part of the optical conductivity of FeTe$_{0.55}$Se$_{0.45}$
 in the infrared region for light polarized in the {\em a-b} planes at several
 temperatures above and below $T_c$.  In the normal state there is a gradual
 shift in spectral weight from high to low frequency.  Below $T_c$ there is
 a dramatic loss of the low-frequency spectral weight.
 Inset: The Drude-Lorentz fit to the data at 100~K.  The Drude
 component (dashed line) and Lorentzian oscillators (long-dashed lines)
 combine (dotted line) to reproduce the data (solid line) quite
 well.}\label{fig:sigma}
\end{figure}

The temperature dependence of the real part of the infrared optical conductivity
is shown in Fig.~\ref{fig:sigma}.  At room temperature the conductivity is flat
and relatively featureless except for an infrared-active $E_u$ mode at 204~cm$^{-1}$
which involves the in-plane displacements of the Fe-Te(Se) atoms \cite{xiatl09}.
As the temperature is lowered there is a shift in the spectral weight from high
to low frequency, where the spectral weight is defined here as the area under
the conductivity curve over a given interval
$$
  N(\omega, T) = \int_0^{\,\omega} \sigma_1(\omega^\prime, T)\, d\omega^\prime .
$$
This is the expected response for a metallic system where the scattering rate
decreases with temperature.  The optical conductivity over most of the temperature
range is described quite well by a simple Drude-Lorentz model for the dielectric
function $\tilde\epsilon = \epsilon_1 +i\epsilon_2$,
$$
  \tilde\epsilon(\omega) = \epsilon_\infty - {{\omega_{p,D}^2}\over{\omega^2+i\omega/\tau_D}}
    + \sum_j {{\Omega_j^2}\over{\omega_j^2 - \omega^2 - i\omega\gamma_j}},
$$
where $\epsilon_\infty$ is the real part of the dielectric function at high
frequency, $\omega_{p,D}^2 = 4\pi ne^2/m^\ast$ and $1/\tau_D$ are the square of
the plasma frequency and scattering rate for the delocalized (Drude) carriers,
respectively; $\omega_j$, $\gamma_j$ and $\Omega_j$ are the position, width,
and strength of the $j$th vibration or excitation.
The complex conductivity is $\tilde\sigma(\omega) = \sigma_1 +i\sigma_2 =
-i\omega [\tilde\epsilon(\omega) - \epsilon_\infty ]/4\pi$.

%
%

\begin{figure}[bthp]
\centerline{\includegraphics[width=7.3cm]{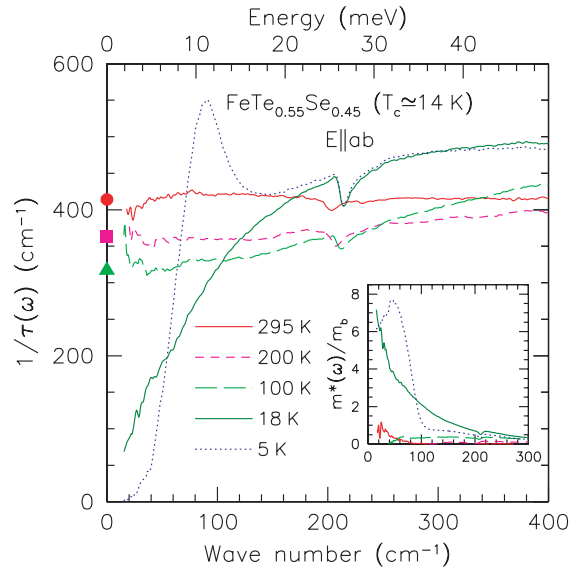}}
\caption{The in-plane frequency-dependent scattering rate of
 FeTe$_{0.55}$Se$_{0.45}$ for several temperatures above and below $T_c$
 in the far-infrared region.  The values for $1/\tau_D$ are shown at 295 ({\CIRCLE}),
 200 ($\blacksquare$) and 100~K ($\blacktriangle$), respectively, where the
 scattering rate displays little temperature dependence.  For $T \gtrsim T_c$
 (18~K) at low frequency $1/\tau(\omega) \propto \omega$, while for $T < T_c$
 large changes in the scattering rate are observed in response to the formation of
 superconducting gap(s).
 Inset: The frequency dependence of the effective mass.}\label{fig:tau}
\end{figure}

The optical conductivity may be reproduced using this model at
295, 200 and 100~K, with fitted values of $\omega_{p,D} = 7200$~cm$^{-1}$ and
$1/\tau_D = 414, 363$ and 317~cm$^{-1}$, respectively ($\pm 5$\%).  To fit the
midinfrared component, Lorentzian oscillators at the somewhat arbitrary
positions of 650 and 3200~cm$^{-1}$ have been introduced; the results of the
fit to the data at 100~K are shown in the inset in Fig.~\ref{fig:sigma}.
An alternative method that has been used in some of the pnictide materials
considers two Drude components \cite{wu10,barisic10}.  If we apply this approach
to the data at 100~K, we note that the ``narrow'' Drude component is similar to
that obtained from the Drude-Lorentz fits, while the ``broad'' Drude component has
a width and strength similar to that of the oscillator at $\simeq 650$~cm$^{-1}$.
While the scattering rates are expected to change with temperature, the plasma
frequencies should remain relatively constant.  However, just above $T_c$ at 18~K
the plasma frequency of the narrow Drude component has unexpectedly decreased by
more than a factor of two.  In addition, at 18~K neither the Drude-Lorentz or
the two-Drude model accurately the shape of the low-frequency conductivity.
To address this problem, we consider the extended-Drude model in which both
the scattering rate and the effective mass take on a frequency dependence, an
approach that has been previously applied to several pnictide materials
\cite{yang09,wu09a}.  The experimentally-determined scattering rate and
effective mass are \cite{allen77,puchkov96}
$$
  {{1}\over{\tau(\omega)}} = {{\omega_p^2}\over{4\pi}} \,
  {\rm Re} \left[ {{1}\over{\tilde\sigma(\omega)}} \right],
$$
and
$$
  {{m^\ast(\omega)}\over{m_b}} = {{\omega_p^2}\over{4\pi\omega}} \,
  {\rm Im} \left[ {{1}\over{\tilde\sigma(\omega)}} \right].
$$
In this instance we set $\omega_p \equiv \omega_{p,D}$ and
$\epsilon_\infty = 4$ (although the choice of $\epsilon_\infty$ has little
effect on the scattering rate or the effective mass in the far-infrared
region).  The temperature dependence of $1/\tau(\omega)$ is shown in
Fig.~\ref{fig:tau}, and the inset shows the temperature dependence of
$m^\ast(\omega)/m_b$.
At 295, 200 and 100~K the scattering rate displays little frequency
dependence, and moreover $1/\tau(\omega\rightarrow 0) \simeq 1/\tau_D$.
This self-consistent behavior indicates that within this temperature range,
the transport may be described as a weakly-interacting Fermi liquid (Drude
model).  However, just above $T_c$ at 18~K the scattering rate develops a
linear frequency dependence $\lesssim 200$~cm$^{-1}$, suggesting the
presence of electronic correlations.  This may be due in
part to magnetic correlations \cite{wen09} that arise from the suppression
of the magnetic transition in Fe$_{1+\delta}$Te at $T_N \simeq 70$~K in
response to Se substitution \cite{sales09}.  We note that similar behavior
of the scattering rate is observed in many optimally-doped cuprate
superconductors where the electronic correlations may have a similar
origin \cite{basov05}.
Dramatic changes are also observed in $1/\tau(\omega)$ below $T_c$ where
the scattering rate is suppressed at low frequencies, but increases rapidly
and overshoots the normal-state (18~K) value at about 60~cm$^{-1}$, finally
merging with the normal-state curve at about 200~cm$^{-1}$; this behavior
is in rough agreement with a recently proposed differential sum rule for the
scattering-rate \cite{basov02,abanov02,chubukov03}.
%
%
%
We note that the overshoot in $1/\tau(\omega)$ below $T_c$ is in general
more characteristic of a material with an {\em s}-wave gap rather than a
higher-order {\em d}-wave gap \cite{marsiglio01}.
In the normal state just above $T_c$, the low-frequency limit for the
effective mass is $m^\ast(\omega\rightarrow 0)/m_b \simeq 7 - 8$,
which is comparable with ARPES estimates of $m^\ast/m_b \simeq 6 - 20$
\cite{tamai10}.  This mass enhancement is too large to be caused by
electron-phonon interactions or coupling to spin-fluctuations alone
\cite{yildirim09}, suggesting that electronic correlations play a
dominant role in the low-energy excitations in this material
\cite{haule08,qazilbash09}.
%
%
%
%
This result is quite different than the behavior of the effective mass
in the sulfide analog Fe$_{1.06}$Te$_{0.88}$S$_{0.14}$ where $m^\ast(\omega)/m_b < 0$
over much of the far-infrared region, from which it is inferred
that the carriers form an ``incoherent metal'' \cite{craco09,aichhorn10}.

\begin{figure}[tb]
\centerline{\includegraphics[width=6.8cm]{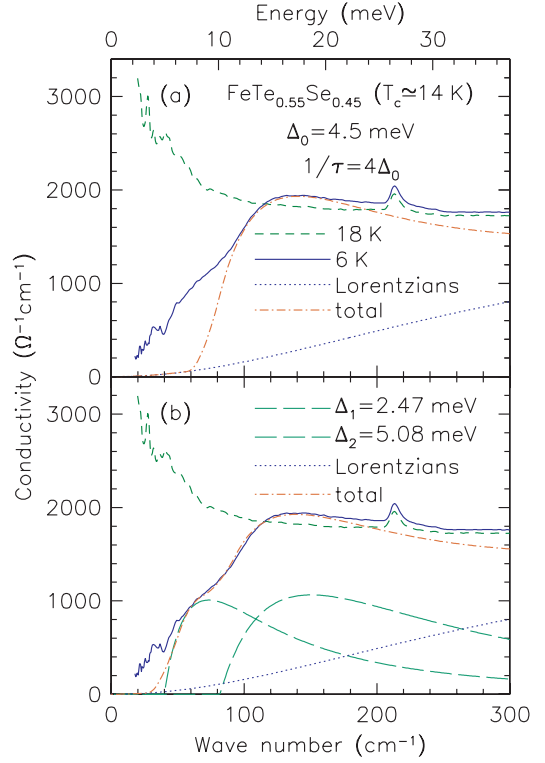}}
\caption{(a) The in-plane optical conductivity of FeTe$_{0.55}$Se$_{0.45}$
shown at 18 and 6~K (dashed and solid lines, respectively).  The optical
conductivity with a single isotropic gap of $2\Delta_0 \simeq 4.5$~meV
with a scattering rate $1/\tau=4\Delta_0$ is calculated for $T \ll T_c$
(long-dashed line) and superimposed on the contribution from the bound
excitations in the mid-infrared (dotted line); the smoothed linear
combination of the two curves (dot-dash line) does not reproduce the
low-frequency data.
(b) The optical conductivity with gaps of $2\Delta_1 \simeq 5$~meV
and $2\Delta_2 \simeq 10.2$~meV  with the scattering rates $1/\tau_j = 4\Delta_j$
is calculated for $T \ll T_c$ and superimposed on the Lorentzian contribution;
the smoothed linear combination of the three curves is in much better agreement
with the measured data below 200~cm$^{-1}$.
}\label{fig:gaps}
\end{figure}

%
%
The low-frequency conductivity is shown in more detail in Fig.~\ref{fig:gaps}.
For $T < T_c$ there is a dramatic suppression of the low-frequency conductivity with
a commensurate loss of spectral weight.  This ``missing area'' is associated with
the formation of a superconducting condensate, whose spectral weight $N_c$ may be
calculated from the Ferrell-Glover-Tinkham sum rule \cite{ferrell58,tinkham59}
$$
  N_c \equiv N(\omega_c, T\simeq T_c) - N(\omega_c, T\ll T_c) = \omega_{p,S}^2/8.
$$
Here $\omega_{p,S}^2 = 4\pi n_s e^2/m^*$ is the square of the superconducting
plasma frequency and superfluid density is $\rho_{s0} \equiv \omega_{p,S}^2$;
the cut-off frequency $\omega_c\simeq 150$~cm$^{-1}$ is chosen so
that the integral converges smoothly.  The superconducting plasma frequency
has also been determined from $\epsilon_1(\omega)$ in the low frequency limit
where $\epsilon_1(\omega) = \epsilon_\infty - \omega_{p,S}^2/\omega^2$.  Yet another
method of extracting $\omega_{p,S}$ from $\epsilon_1(\omega)$ is to determine
$[-\omega^2\epsilon_1(\omega)]^{1/2}$ in the $\omega\rightarrow 0$ limit \cite{jiang96}.
All three techniques yield $\omega_{p,S} = 3000\pm 200$~cm$^{-1}$, indicating that less
than one-fifth of the free-carriers in the normal state have condensed
($\omega_{p,S}^2/\omega_{p,D}^2 \lesssim 0.18$), implying that this material
is not in the clean limit.
The superfluid density can also be expressed as an effective penetration
depth $\lambda_0 = 5300\pm 300$~\AA , which is in good agreement with recent
tunnel-diode \cite{kim10} and muon-spin spectroscopy \cite{biswas10} measurements
on FeTe$_{0.63}$Se$_{0.37}$ and FeTe$_{0.5}$Se$_{0.5}$, respectively.

%
%
%
The strong suppression of the conductivity for $T < T_c$ below
$\simeq 120$~cm$^{-1}$ is characteristic of the opening of a superconducting
energy gap in the density of states at the Fermi surface; in addition, there
also appears to be a shoulder at $\simeq 60$~cm$^{-1}$.  As previously noted,
the formation of a gap leads to a transfer of spectral weight into the condensate.
Below $T_c$, the optical conductivity has been calculated using a Mattis-Bardeen
approach \cite{mattis58} for the contribution from the gapped excitations
\cite{zimmerman91}. This method assumes that $l \lesssim \xi_0$, where the
mean-free path $l=v_{\rm F}\tau$ ($v_{\rm F}$ is the Fermi velocity), and the coherence
length $\xi_0 = \hbar v_{\rm F}/\pi\Delta_0$ for an isotropic superconducting
energy gap $\Delta_0$; this may also be expressed as $1/\tau \gtrsim 2\Delta_0$.
The dirty-limit approach is consistent with the observation that less than 20\%
of the free carriers collapse into the condensate.  Initially, only a single
isotropic gap $\Delta_0 \simeq 4.5$~meV for $T\ll T_c$ was considered; however,
even with a moderate amount disorder scattering ($1/\tau = 4\Delta_0$) and the
addition of the low-frequency tail from the bound Lorentzian oscillators this fails
to accurately reproduce the low-frequency conductivity, as Fig.~\ref{fig:gaps}(a)
demonstrates.  To properly model the conductivity two gap features have been
considered, $\Delta_1 \simeq 2.5$~meV and $\Delta_2 \simeq 5.1$~meV with
$1/\tau_j = 4\Delta_j$ for $T\ll T_c$.  The observation of two gap features
is consistent with recent optical \cite{kim09,heumen09,perucchi10}, ARPES \cite{ding08},
microwave \cite{hashimoto09} and penetration depth \cite{malone09} results
on the pnictide compounds, as
well as some theoretical works that propose that {\em s}-wave gaps form on
the hole ($\alpha$) and electron ($\beta$) pockets, possibly with a sign
change between them \cite{mazin08,kuroki08,chubukov08}, the so-called
$s^\pm$ symmetry state.  In the $s^\pm$ model, the gap on the electron
pocket ($\beta$) may be an extended {\em s}-wave and have nodes on its
Fermi surface \cite{chubukov09}. While there is some uncertainty associated
with the low-frequency conductivity in this work, the apparent lack
of residual conductivity in the terahertz region for $T\ll T_c$ suggests the
absence of nodes.  It is possible that disorder may lift the nodes, resulting
in a nodeless extended {\em s}-wave gap \cite{mishra09,carbotte10}.  While
the optical gaps provide estimates of the gap amplitudes, they do not
distinguish between $s^\pm$ and extended {\em s}-wave.
The optical gaps at $2\Delta_j \simeq 40$ and 82~cm$^{-1}$ are either
similar to or slightly larger than the low-frequency scattering rate
observed at 18~K, $1/\tau(\omega\rightarrow 0) \simeq 40$~cm$^{-1}$.
This might seem to suggest that the Mattis-Bardeen approach should not
be used;  however, the linear frequency dependence of the scattering rate
complicates matters.  If we consider the value of the scattering rate in
the region of the optical gaps where the scattering should be important,
then from Fig.~\ref{fig:tau} we have

$$
  {{1/\tau_j(2\Delta_j)}\over{2\Delta_j}} \simeq 3
$$
which is actually larger than the ratio of 2 that was assumed in the calculation,
indicating that the Mattis-Bardeen approach is valid.  While the value of
$2\Delta_1/k_{\rm B}T_c \simeq 4$ is close the value of 3.5 expected for a BCS
superconductor in the weak-coupling limit, $2\Delta_2/k_{\rm B}T_c \simeq 8.4$
is significantly larger.

\begin{figure}[t]
\centerline{\includegraphics[width=7.3cm]{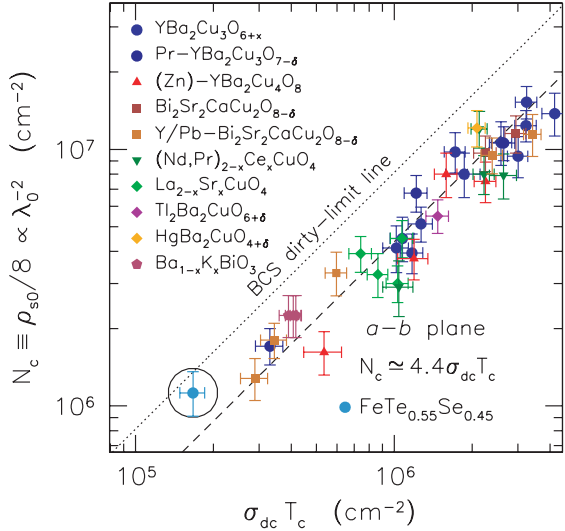}}
\caption{The log-log plot of the in-plane spectral weight of the superfluid
 density $N_c \equiv \rho_{s0}/8$ vs $\sigma_{dc}\,T_c$, for a variety of
 electron and hole-doped cuprates compared with the result for
 FeTe$_{0.55}$Se$_{0.45}$.  The dashed line corresponds to the general
 result for the cuprates $\rho_{s0}/8 \simeq 4.4 \sigma_{dc}T_c$, while
 the dotted line is the result expected for a BCS dirty-limit supercondcutor
 in the weak-coupling limit, $\rho_{s0}/8 \simeq 8.1\, \sigma_{dc}T_c$.
}\label{fig:scaling}
\end{figure}

%
%
It was recently noted that in a number of the pnictide materials \cite{wu09b}
the superfluid density $\rho_{s0}$ falls on a recently proposed empirical
scaling relation for the cuprate superconductors shown by the dashed line
in Fig.~\ref{fig:scaling} \cite{homes04,homes05},
$$
 \rho_{s0}/8 \simeq 4.4\, \sigma_{dc}\, T_c .
$$
From the estimate of $\sigma_{dc} \equiv \sigma_1(\omega \rightarrow 0) =
3500\pm 400$~$\Omega^{-1}$cm$^{-1}$ for $T \gtrsim T_c$ (determined from
Fig.~\ref{fig:sigma}, as well as Drude-Lorentz fits), and the previously
determined value of $\rho_{s0} = 9\pm 1 \times 10^6$~cm$^{-2}$,
we can see that FeTe$_{0.55}$Se$_{0.45}$ also falls close to this scaling line.
In fact, in a BCS dirty-limit superconductor in the weak-coupling limit, the numerical
constant in the scaling relation is calculated to be slightly larger \cite{homes05}
$$
 \rho_{s0}/8 \simeq 8.1\, \sigma_{dc}\, T_c
$$
(dotted line in Fig.~\ref{fig:scaling}); the result for this material is
actually closer to the BCS dirty-limit line than the one established for the
cuprate superconductors.

%
%

\section{Conclusion}
To summarize, the optical properties of FeTe$_{0.55}$Se$_{0.45}$ ($T_c = 14$~K)
have been examined for light polarized in the Fe-Te(Se) planes above and below
$T_c$.  Well above $T_c$ the transport may be described by a weakly-interacting
Fermi liquid (Drude model); however, this picture breaks down close to $T_c$
when the scattering rate takes on a linear frequency dependence, similar to
what is observed in the cuprate superconductors.
Below $T_c$, less than one-fifth of the free carriers collapse into the
condensate ($\lambda_0 \simeq 5300$~\AA ), indicating that this material is
in the dirty limit, and indeed this material falls on the general scaling
line predicted for a BCS dirty-limit superconductor in the weak coupling limit.
To successfully model the optical conductivity in the superconducting state,
two gaps of $\Delta_1 \simeq 2.5$~meV and $\Delta_2 \simeq 5.1$~meV are considered
using a Mattis-Bardeen formalism (with moderate disorder scattering), suggesting
either an $s^\pm$ or a nodeless extended {\em s}-wave gap.

%
%
We would like to acknowledge useful discussions with D. N. Basov,
J. P. Carbotte, A. V. Chubukov, S. V. Dordevic, D. C. Johnson, D. J. Singh,
J. M. Tranquada, and J. J. Tu.
JSW and ZJX are supported by the Center for Emergent Superconductivity,
an Energy Frontier Research Center funded by the U.S. Department of Energy,
Office of Basic Energy Sciences.  This work is supported by the Office of
Science, U.S. DOE under Contract No. DE-AC02-98CH10886.



%
%

%

%
\end{document}